\documentclass[conference]{IEEEtran}
\IEEEoverridecommandlockouts
\usepackage{cite}
\usepackage{amsmath,graphicx}
\usepackage{amssymb,amsfonts}
\usepackage{graphicx}
\usepackage{textcomp}
\usepackage{xcolor}
\usepackage{subfigure}
\usepackage{textcomp} % to use \textperthousand                                                     
\usepackage{siunitx}  
\usepackage{algorithm}
\usepackage{algpseudocode}
\usepackage{xpatch}
\usepackage{pgfplots}
\usepackage{pgfplotstable}
\usepackage{readarray}
\usepackage{siunitx}
\usepackage{bm}
\usepackage[nolist]{acronym}
\usepackage{bbm}
\usepackage{url}

\usetikzlibrary{spy}

\DeclareMathOperator{\diag}{diag}
\DeclareMathOperator{\He}{H}

\newcommand{\eyeM}{\bm{\mathrm{I}}}
\newcommand{\bbthet}{\bm{\bar{\theta}}}
\newcommand{\bD}{\bm{D}}
\newcommand{\bF}{\bm{F}}
\newcommand{\Her}{{\mathrm{H}}}

\newcommand{\bC}{\bm{C}}

\newcommand{\NB}{M}%{N_{\text{B}}}
\newcommand{\NM}{K}%{N_{\text{M}}}
\newcommand{\NRe}{N}%{N_{\text{R}}}

\newcommand{\mThet}{\bm{\Theta}}

\newcommand{\bv}{\bm{v}}

\newcommand{\norm}[1]{\left\|#1\right\|}

\newcommand{\abs}[1]{\left|#1\right|}

\newcommand{\summe}[2]{\sum_{#1}^{#2}}

\newcommand{\teR}{{\text{R}}}

\newcommand{\teRS}{{\text{RS}}}
\newcommand{\teDS}{{\text{DS}}}
\newcommand{\teDR}{{\text{DR}}}

\newcommand{\teDN}{{\text{DeN}}}

\newcommand{\bi}{\bm{i}}
\newcommand{\bb}{\bm{b}}
\newcommand{\be}{\bm{e}}

\newcommand{\ba}{\bm{a}}

\newcommand{\bx}{\bm{x}}

\newcommand{\bZ}{\bm{Z}}
\newcommand{\bA}{\bm{A}}
\newcommand{\bz}{\bm{z}}

\newcommand{\bzero}{\bm{0}}
\newcommand{\tG}{{\text{G}}}
\newcommand{\teL}{{\text{L}}}
\newcommand{\tN}{{\text{N}}}

\newcommand{\inv}{{-1}}
\newcommand{\transpo}{{\mathrm{T}}}
\newcommand{\im}{\mathrm{j}}

\newcommand{\Real}[1]{\mathrm{Re}{#1}}
\newcommand{\Imag}[1]{\mathrm{Im}{#1}}
\newcommand{\eye}{\bm{1}}

\definecolor{TUMBeamerYellow}    {rgb} {1.000,0.706,0.000}    % RGB 255,180,000
\definecolor{TUMBeamerOrange}    {rgb} {1.000,0.502,0.000}    % RGB 255,128,000
\definecolor{TUMBeamerRed}       {rgb} {0.898,0.204,0.094}    % RGB 229,052,024
\definecolor{TUMBeamerDarkRed}   {rgb} {0.792,0.129,0.247}    % RGB 202,033,063
\definecolor{TUMBeamerBlue}      {rgb} {0.000,0.600,1.000}    % RGB 000,153,255
\definecolor{TUMBeamerLightBlue} {rgb} {0.255,0.745,1.000}    % RGB 065,190,255
\definecolor{TUMBeamerGreen}     {rgb} {0.569,0.675,0.420}    % RGB 145,172,107
\definecolor{TUMBeamerLightGreen}{rgb} {0.710,0.792,0.510}    % RGB 181,202,130

\begin{acronym}
    \acro{AoD}{angle of departure}
    \acro{AoA}{angle of arrival}
    \acro{ULA}{uniform linear array}
    \acro{CSI}{channel state information}
    \acro{LOS}{line of sight}
    \acro{EVD}{eigenvalue decomposition}
    \acro{BS}{base station}
    \acro{MS}{mobile station}
    \acro{mmWave}{millimeter wave}
    \acro{DPC}{dirty paper coding}
    \acro{RIS}{reconfigurable intelligent surface}
    \acro{AWGN}{additive white gaussian noise}
    \acro{MIMO}{multiple-input multiple-output}
    \acro{SISO}{single-input single-output}
    \acro{UL}{uplink}
    \acro{DL}{downlink}
    \acro{OFDM}{orthogonal frequency-division multiplexing}
    \acro{TDD}{time-division duplex}
    \acro{LS}{least squares}
    \acro{MMSE}{minimum mean square error}
    \acro{SINR}{signal to interference plus noise ratio}
    \acro{OBP}{optimal bilinear precoder}
    \acro{LMMSE}{linear minimum mean square error}
    \acro{MRT}{maximum ratio transmitting}
    \acro{M-OBP}{multi-cell optimal bilinear precoder}
    \acro{S-OBP}{single-cell optimal bilinear precoder}
    \acro{SNR}{signal-to-noise ratio}
    \acro{SDR}{Semidefinite Relaxation}
    \acro{SE}{spectral efficiency}
    \acro{GCEs}{Gram channel eigenvalues}
    \acro{LNA}{low-noise amplifier}
    \acro{Tx}{transmit side}
    \acro{Rx}{receive side}
    \acro{HPA}{high power amplifier}
    \acro{LNA}{low noise amplifier}
\end{acronym}

\def\BibTeX{{\rm B\kern-.05em{\sc i\kern-.025em b}\kern-.08em
    T\kern-.1667em\lower.7ex\hbox{E}\kern-.125emX}}
\begin{document}

\title{Performance Analysis of Systems\\ with Coupled and Decoupled RISs
}

\author{\IEEEauthorblockN{Dominik Semmler, Josef A. Nossek, \textit{Life Fellow, IEEE}, Michael Joham, and Wolfgang Utschick, \textit{Fellow, IEEE}}
\IEEEauthorblockA{\textit{School of Computation, Information and Technology, Technical University of Munich, 80333 Munich, Germany} \\
email: \{dominik.semmler,josef.a.nossek,joham,utschick\}@tum.de}}

\maketitle

\begin{abstract}
We analyze and compare different methods for handling the mutual coupling in \ac{RIS}-aided communication systems.
A new mutual coupling aware algorithm is derived where the reactance of each element is updated successively with a closed-form solution.
In comparison to existing element-wise methods, this approach leads to a considerably reduced computational complexity.
Furthermore, we introduce decoupling networks for the \ac{RIS} array as a potential solution for handling mutual coupling.
With these networks, the system model reduces to the same structure as when no mutual coupling were present.
Including decoupling networks, we can optimize the channel gain of a \ac{RIS}-aided \ac{SISO} system in closed-form which allows to analyze the scenario under mutual coupling analytically and to draw connections to the conventional transmit array gain.
In particular, a super-quadratic channel gain can be achieved which scales as $ \NRe^4$ where $\NRe$ is the number of \ac{RIS} elements.
\end{abstract}


\begin{figure}[b]
    \onecolumn
    \centering
    \scriptsize{This work has been submitted to the IEEE for possible publication. Copyright may be transferred without notice, after which this version may no longer be accessible.}
    \vspace{-1.3cm}
    \twocolumn
\end{figure}

\section{Introduction}
\Acp{RIS} are currently highly discussed as they are considered an important technology for future wireless communications systems (see \cite{Power_Min_IRS}, \cite{SmartRadioEnvironment}).
\Acp{RIS} are arrays consisting of many passive reflecting elements which can manipulate the incoming wavefronts and, hence, shape the propagation environment.
The potential of \acp{RIS} has already been demonstrated and including a \ac{RIS} has shown to significantly improve the performance in various scenarios, e.g., the power consumption \cite{Power_Min_IRS} or the energy efficiency \cite{EnergyEff}.
In these publications and in the majority of \ac{RIS} literature, a simple model is used where each of the reflecting elements is modeled by a phase manipulation.

More practical models for \acp{RIS} have already been introduced in the literature.
For example in \cite{MutualCouplingAware}, a \ac{RIS} model was introduced that is based on impedance matrices.
Impedance-based descriptions have already shown to provide a powerful description for conventional communication systems without the \ac{RIS} (see \cite{TowardCircuitTheory}).
Nevertheless, the same electromagnetic system can be described in various different, equally valid, ways where many parametrizations are possible.
For example in \cite{ScatteringRIS}, the \ac{RIS} has been modeled based on scattering parameters and with the correct interpretation of the matrices (see \cite{NewChannelModel}), both approaches lead to the same conclusions.
In this article, only impedance-based descriptions according to \cite{MutualCouplingAware} are used as they lead to more convenient expressions for our purposes.

\begin{figure}
    \includegraphics{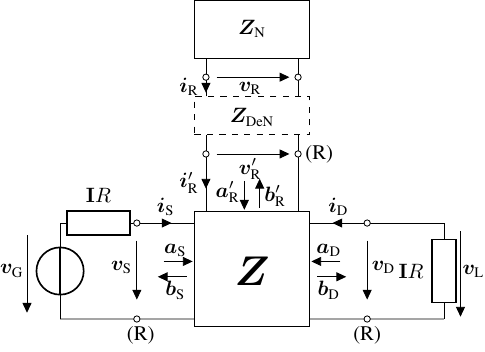}
    \caption{Multiport Model of \cite{NewChannelModel} with an additional decoupling network $\bZ_\teDN$}
    \label{fig:Three_Port}
\end{figure}

With the more practical descriptions, as mentioned above, it is possible to include the mutual coupling between the reflecting elements in the system model.
Modeling the mutual coupling is generally recommended as the reflecting elements typically have a small distance to each other.
In \cite{MutualCouplingZAlgo}, mutual coupling was analyzed based on impedance matrices and an efficient algorithm by using the Neumann series has been designed for the \ac{SISO} channel gain maximization.
The impact of mutual coupling on the rate has been further investigated in \cite{MutualCouplingSumRate} and \cite{MutualCouplingGradient} whereas in \cite{FollowUpBDRIS} a mutual coupling aware algorithm for beyond diagonal \acp{RIS} was designed.
Furthermore, in \cite{FollowUpSZ} and \cite{FollowUPSZJournal} two mutual coupling aware algorithms based on the scattering parameters have been proposed which show better performance than the one of \cite{MutualCouplingZAlgo}.
These algorithms rely on the Neumann series approximation and in \cite{ElementWise}, an element-wise algorithm was derived in which no approximation is needed and where each reactance is computed in closed-form.
However, this algorithm has a complexity of $\mathcal{O}(\NRe^4)$ which is one order of magnitude higher than for the other algorithms.

In this article, we give a different approach on the element-wise method which has only a complexity of  $\mathcal{O}(\NRe^3)$.
Additionally, we introduce decoupling networks as a potential solution to handle the mutual coupling of the \ac{RIS} array.
Decoupling networks have already been used for the transmit and receive arrays (see \cite{TowardCircuitTheory}) but we will show that there are significant advantages when also including them for the \ac{RIS} array.
In particular, we make the following contributions in this article:
\begin{itemize}
    \item We derive a new approach for the element-wise algorithm which reduces the complexity by one order of magnitude.
    \item Decoupling networks are introduced and compared to mutual coupling aware algorithms.
          Including these networks has the advantage that all conventional solutions, neglecting the mutual coupling, can be directly extended to the case of mutual coupling with this approach.
    \item By using decoupling networks, we can optimize the channel gain of a \ac{RIS}-aided \ac{SISO} system in closed-form which allows us to analytically analyze the array gain of a \ac{RIS} under mutual coupling.
          From this analysis, we can draw connections to the conventional transmit array gain.
          For example, we show that a super-quadratic array gain which scales as $\NRe^4$ is possible in a \ac{RIS}-aided scenario.
\end{itemize}

\section{System Model}
We consider a \ac{RIS}-aided point-to-point \ac{MIMO} system where one \ac{BS}, having $\NB$ antennas, transmits data signals to one user, having $\NM$ antennas.
The transmission is supported by a \ac{RIS}, consisting of $\NRe$ reflecting elements. By using
the impedance representation, we arrive at the channel from the \ac{BS} to the user (see \cite{MutualCouplingAware}, \cite{NewChannelModel})
\begin{equation}\label{eq:ChannelModel}
    \begin{aligned}
    \bv_\teL &= \bD \bv_\tG = \frac{1}{4R} \bZ \bv_\tG\quad \text{with}\\
    \bZ &=  \bZ_\teDS - \bZ_\teDR( \bZ_\teR+ \bZ_\tN )^{\inv}\bZ_\teRS\\
    \end{aligned}
\end{equation}
with the definition of $\bv_\tG$ and $\bv_\teL$ according to Fig. \ref{fig:Three_Port}.
Here, $\bZ_\teDS$ is the direct channel from the \ac{BS} to the user, $\bZ_\teDR$ is the channel from the \ac{RIS} to the user, and $\bZ_\teRS$ the channel from the \ac{BS} to the \ac{RIS}.
The adjustable passive impedance network of the \ac{RIS} is given by $\bZ_\tN$,  whereas $\bZ_\teR$ is the impedance matrix of the \ac{RIS} antenna array accounting for mutual coupling.
If there were no mutual coupling present, $\bZ_\teR = \eyeM R$ would be a diagonal matrix.
In this article, we analyze the aspects of mutual coupling.
For isotropic radiators (which we consider in this article), the matrix $\bZ_\teR$ is defined as (see \cite[Eqn. (5)]{NewChannelModel}) 
\begin{equation}
    [\bZ_\teR]_{i,j} = R \frac{\sin(2\pi \abs{i-j})}{2\pi \abs{i-j}} + \im R \frac{\cos(2\pi \abs{i-j})}{2\pi \abs{i-j}} \quad \text{for} \; i\neq j,
\end{equation}
 whereas the diagonal entries are given by  $[\bZ_\teR]_{i,i} = R$.
Additionally, we assume a single-connected \ac{RIS} and the adjustable impedance network to be lossless and, hence,
the matrix $\bZ_\tN$ is a diagonal purely imaginary matrix 
\begin{equation}\label{eq:AdjNetwork}
    \bZ_\tN = \im\diag(\bx)
\end{equation}
with $\bx=[x_1, x_2, \dots, x_N]^\transpo \in \mathbb{R}^N$ being purely real-valued. 

\section{Element-Wise Algorithm}
In \cite{ElementWise}, an algorithm was proposed which successively updates the reactances in closed-form.
However, its complexity scales as $\mathcal{O}(\NRe^4)$.
We will give a different approach to an element-wise algorithm which reduces the complexity to  $\mathcal{O}(\NRe^3)$.
It is important to note that this approach can be extended to more practical considerations (e.g. to a \ac{RIS} with losses) as was also analyzed in \cite{ElementWise}.
We will, however, only consider ideal assumptions in this article.
When updating the $n$-th reactance,  we can rewrite its value by
\begin{equation}\label{eq:EWUpdatem}
    x_n =  \bar{x}_n + {\Delta x}_n
\end{equation}
where $\bar{x}_n$ is the old value of the $n$-th reactance and ${\Delta x}_n$ is the update which we would like to determine in closed-form.
The other reactances $x_m, \; m\neq n$, stay constant which is highlighted by referring to them also as $\bar{x}_m$.
Inserting \eqref{eq:EWUpdatem} and \eqref{eq:AdjNetwork} into \eqref{eq:ChannelModel}, we can write the channel matrix as 
\begin{equation}\label{eq:EWChannelModelUpdatem}
    \begin{aligned}
        \bZ 
         &= \bZ_\teDS - \bZ_\teDR(\bZ_\teR + \im \diag(\bm{\bar{x}}) +\im {\Delta x}_n \be_n \be_n^\transpo )^{\inv}\bZ_\teRS
    \end{aligned}
\end{equation}
where $\be_n$ is the $n$-th standard basis vector. 
Applying the matrix inversion lemma, the inverse in Eqn. \eqref{eq:EWChannelModelUpdatem} can be reformulated as
\begin{equation}\label{eq:EWInverseROUpdate}
    \begin{aligned}
        \bZ^{-1}_{\text{RIS}} &=(\bZ_\teR + \im \diag(\bm{\bar{x}}) + \im {\Delta x}_n \be_n \be_n^\transpo)^{\inv} \\
        &=  \bm{\bar{Z}}^{-1}_{\text{RIS}} -  \bm{\bar{Z}}^{-1}_{\text{RIS}}\be_n \be_n^\transpo \bm{\bar{Z}}^{-1}_{\text{RIS}}  \frac{\im {\Delta x}_n}{1+ \im {\Delta x}_n \be_n^\transpo \bm{\bar{Z}}^{-1}_{\text{RIS}}\be_n}
    \end{aligned}
\end{equation}
where $\bm{\bar{Z}}^{-1}_{\text{RIS}} = (\bZ_\teR + \im \diag(\bm{\bar{x}}))^{\inv}$. 
This allows us to further simplify the channel expression in \eqref{eq:EWChannelModelUpdatem} as
\begin{equation}\label{eq:EWRankOneUpdate}
    \bZ = \bm{\bar{Z}} + \ba \bb^{\prime,\He}\frac{\im {\Delta x}_n}{1+ \im {\Delta x}_n g}
\end{equation}
with the parameters
\begin{gather}
    \begin{aligned}
        \bm{\bar{Z}} &=\bZ_\teDS - \bZ_\teDR\bm{\bar{Z}}^{-1}_{\text{RIS}} \bZ_\teRS,    \\
      \ba &=\bZ_\teDR\bm{\bar{Z}}^{-1}_{\text{RIS}}\be_n, \quad \bb^{\prime,\Her}= \be_n^\transpo \bm{\bar{Z}}^{-1}_{\text{RIS}}\bZ_\teRS,\quad g= \be_n^\transpo \bm{\bar{Z}}^{-1}_{\text{RIS}}\be_n.
      \raisetag{20pt}
    \end{aligned}
\end{gather}

\subsection{Change of Optimization Parameters}
To obtain more convenient expressions in the following derivations, we conduct a change of optimization parameters which has been introduced in \cite{MutualCouplingZAlgo}.
For a complex number $z$, it is possible to get the following expressions (cf. \cite[Eqns. (6), (7)]{MutualCouplingZAlgo})
\begin{equation}\label{eq:ChangeOfOptParameters}
    z = \frac{2\Real{(z)}}{ 1+e^{\im \phi}}, \quad  \Imag{(z)} = -\Real{(z)} \tan(\phi/2) \quad \text{for} \;\Real{(z)} \neq 0
\end{equation}
in which the optimization variable $x = \Imag{z}   \in \mathbb{R}$ is converted to an angle $\phi \in [-\pi,\pi]$.
It is important to note that this change of optimization parameters is purely for mathematical convenience and for numerical considerations.
The approaches are equivalent and can be converted to each other.
In particular, we use the following change of optimization parameters
\begin{equation}
    \frac{ \im {\Delta x}_n}{1+ g\im {\Delta x}_n }  = \frac{ 1}{ \im \frac{-1}{{\Delta x}_n} +g} = \frac{1+\theta}{ 2\mathrm{Re}(g)}
\end{equation}
with the new optimization variable $\theta = e^{\im \phi}$ according to equation \eqref{eq:ChangeOfOptParameters}.
The reactance ${\Delta x}_n$ can always be recovered
\begin{equation}
    {\Delta x}_n = \frac{1}{ \mathrm{Re}(g)\tan(\arg(\theta)/2)+\mathrm{Im}(g)}.
\end{equation}
With this change of variables, Eqn. \eqref{eq:EWRankOneUpdate} can be rewritten as 
\begin{equation}\label{eq:EWRankOneUpdateTheta}
    \bZ = \bZ_0 + \ba \bb^{\He}\theta
\end{equation}
\begin{equation}
    \text{with} \quad \bZ_0 = \bm{\bar{Z}} +  \ba \bb^{\He}, \quad \bb^{\He} = \frac{1}{2\Real{(g)}} \bb^{\prime,\He}.
\end{equation}
Due to $\abs{\theta} = 1$, the Gram matrix of $\bZ$ is given by 
\begin{equation}\label{eq:GramChannelMatrix}
    \begin{aligned}
        \bZ \bZ^\Her &=  \bZ_0\bZ_0^\Her + \bZ_0 \bb \ba^\Her \theta^* + \ba \bb^\Her \bZ_0^\Her \theta + \ba \ba^\Her \norm{\bb}_2^2 \\
        &=  \bZ_0\left(\eyeM - \frac{\bb \bb^\Her}{\norm{\bb}_2^2}\right)\bZ_0^\Her + \bF \bbthet \bbthet^\Her \bF^\Her
    \end{aligned}
\end{equation}
with the matrix $\bF$ and the vector $\bbthet$ defined as
\begin{equation}\label{eq:GramRankOneParameters}
        \bF = \left[\ba \norm{\bb}_2, \bZ_0  \frac{\bb}{\norm{\bb}_2}\right], \quad \bbthet^\Her = [\theta^*,1].
\end{equation}
\subsection{SISO Channel Gain}
In this article, we mainly focus on the maximization of the channel gain in a \ac{RIS}-aided \ac{SISO} system.
Here, $\NB=\NM=1$ holds and according to \eqref{eq:EWRankOneUpdateTheta}, the channel gain can be written as 
\begin{equation}
    \abs{z}^2  = \abs{z_0 + a b^*\theta}^2
\end{equation}
Hence, for this special case, the closed-form solution reads as
\begin{equation}
    \phi^{\star} = \arg(z_0)+\arg(b)-\arg(a).
\end{equation}
\subsection{Complexity}
In comparison to \cite{ElementWise}, the calculation in \eqref{eq:EWInverseROUpdate} can be successively computed with the matrix inversion lemma.
The new matrix $\bm{{Z}}^{-1}_{\text{RIS}}$ is updated based on the old matrix $\bm{\bar{Z}}^{-1}_{\text{RIS}}$ with only a rank-one update and, hence, no explicit matrix inversion is needed in each step.
For updating one reactance, the complexity w.r.t. $\NRe$ is therefore $\mathcal{O}(\NRe^2)$.
It follows, that the complexity of updating all elements once is $\mathcal{O}(\NRe^3)$.
\subsection{Extensions}
With the formulation of the Gram matrix in \eqref{eq:GramChannelMatrix}, the algorithm can be extended to various objective functions.
As an example, the \ac{SE} (as considered in \cite{ElementWise},\cite{MutualCouplingGradient}), can be reformulated as
\begin{equation}
    \begin{aligned}
        \log_2\det\left( \eyeM + \bZ \bZ^\Her\right) &= \log_2\det\left( \bA + \bF  \bbthet \bbthet^\Her \bF^\Her\right)\\
        &=\log_2\det( \bA)+ \log_2(1  +   \bbthet^\Her \bC \bbthet)
    \end{aligned}
\end{equation}
where $\bA =  \eyeM + \bZ_0\left(\eyeM - \frac{\bb \bb^\Her}{\norm{\bb}_2^2}\right)\bZ_0^\Her$ and $\bC = \bF^\Her\bA^{\inv}\bF$.
Rewriting the quadratic form as
\begin{equation}
    \begin{aligned}
        \bbthet^\Her \bC  \bbthet 
        &= c_{11}+ c_{22} + 2 \Real{(c_{12}\theta^*) }, \\
    \end{aligned}
\end{equation}
the closed-form solution for the $n$-th reactance reads as
\begin{equation}
    \theta^{\star} = \frac{c_{12}}{\abs{c_{12}}}
\end{equation}
because $c_{11}$ and $c_{22}$ are non-negative real numbers.

\section{Decoupling Networks}
In the last section, we derived an algorithm which is able to take into account the mutual coupling of the \ac{RIS} array.
However, the cost per iteration is $\mathcal{O}(\NRe^3)$ and existing algorithms have a similar complexity.
Additionally, the optimization is numerically challenging and mutual coupling aware algorithms are not guaranteed to find the global optimum.
As a solution, we propose decoupling networks which are well studied for the transmit and receive arrays (see \cite{TowardCircuitTheory}).
These networks can also be applied at the \ac{RIS} antenna array (see Fig. \ref{fig:Three_Port}) and their performance will be analyzed in the following.

\subsection{New Cannnel Model}
In this section, we derive the new channel model when incorporating a decoupling network $\bZ_\teDN$ for the \ac{RIS} array according to Fig. \ref{fig:Three_Port}.
Please note that $\bZ_\teDN = \bZ_\teDN^\transpo$ and $\Real{ (\bZ_\teDN) }= \bzero$ holds as the decoupling network is assumed to be reciprocal and lossless.
Without the decoupling network, the configurable impedance network can be described by the equation
\begin{equation}\label{eq:ConvDecouplingNetwork}
    \bv_\teR = - \bZ_\tN \bi_\teR.
\end{equation}
However, when including the decoupling network according to Fig. \ref{fig:Three_Port}, we have the new relationship 
\begin{equation}\label{eq:NewZN}
    \bv^\prime_\teR = - \bZ_\tN^\prime \bi^\prime_\teR.
\end{equation}
In the following, we derive $\bZ_\tN^\prime$.
With the decoupling network $\bZ_\teDN$, we obtain
\begin{equation}\label{eq:DecouplingNetworkGeneral}
    \begin{bmatrix}
        \bv_\teR\\
        \bv_\teR^\prime
    \end{bmatrix}
    = 
    \begin{bmatrix}
         \bZ_{\teDN,11}& \bZ_{\teDN,12}\\
         \bZ_{\teDN,12}^\transpo& \bZ_{\teDN,22}
    \end{bmatrix}
\begin{bmatrix}
    \bi_\teR\\
    -\bi_\teR^\prime
\end{bmatrix}
\end{equation}
with the four $\NRe \times \NRe$ matrix blocks $\bZ_{\teDN,ij}$. 
Substituting \eqref{eq:ConvDecouplingNetwork} into the first line of \eqref{eq:DecouplingNetworkGeneral}, we obtain 
\begin{equation}
    \bi_\teR =  (\bZ_{\teDN,11} +  \bZ_\tN)^{\inv}\bZ_{\teDN,12}\bi_\teR^\prime.
\end{equation}
Together with the second line of \eqref{eq:DecouplingNetworkGeneral}, we arrive at
\begin{equation}
    \bv_\teR^\prime = - (\bZ_{\teDN,22}-\bZ_{\teDN,12}^\transpo(\bZ_{\teDN,11} +  \bZ_\tN)^{\inv}\bZ_{\teDN,12})\bi_\teR^\prime.
\end{equation}
Therefore, we have found the new matrix [cf. Eqn. \eqref{eq:NewZN}]
\begin{equation}
    \bZ_\tN^\prime = \bZ_{\teDN,22}-\bZ_{\teDN,12}^\transpo(\bZ_{\teDN,11} +  \bZ_\tN)^{\inv}\bZ_{\teDN,12}
\end{equation}
with the new channel model [cf. Eqn \eqref{eq:ChannelModel}]
\begin{equation}\label{eq:DecouplingChannelModel}
    \bZ^\prime =  \bZ_\teDS - \bZ_\teDR( \bZ_\teR+  \bZ_\tN^\prime)^{\inv}\bZ_\teRS.\\
\end{equation}

With the decoupling network $\bZ_{\teDN}$, we can now manipulate the expression for the channel in \eqref{eq:DecouplingChannelModel}.
Various choices exist for decoupling networks, see \cite{TowardCircuitTheory}.
However, in this article, we focus on power matching networks and choose (see \cite[Eqn. (103)]{TowardCircuitTheory})
\begin{equation}
    \bZ_{\teDN} = -\im \begin{bmatrix}
        \bzero& \sqrt{R} \Real{(\bZ_\teR)}^{\frac{1}{2}}\\
        \sqrt{R} \Real{(\bZ_\teR)}^{\frac{1}{2}}&  \Imag{(\bZ_\teR)}
    \end{bmatrix}
\end{equation}
with $ \Real{(\bZ_\teR)} = \Real{(\bZ_\teR)}^{\frac{1}{2}} \Real{(\bZ_\teR)}^{\frac{1}{2}}$ because this particular network structure allows to completely decouple the \ac{RIS} array, where the same structure is obtained as when no mutual coupling were present.
This will be shown in the following.
By using the power matching network, we arrive at
\begin{equation}
    \begin{aligned}
    \bZ_\teR + \bZ_\tN^\prime &=\bZ_\teR - \im \Imag{(\bZ_\teR)}+R\Real{(\bZ_\teR)}^{\frac{1}{2}}\bZ_\tN^{\inv}\Real{(\bZ_\teR)}^{\frac{1}{2}}\\
    &=\Real{(\bZ_\teR)}^{\frac{1}{2}} (\eyeM +R\bZ_\tN^{\inv}) \Real{(\bZ_\teR)}^{\frac{1}{2}}.\\
    \end{aligned}
\end{equation}
As we are considering a single-connected \ac{RIS}, we have $\bZ_\tN=\im \diag(\bx)$ and its inverse $\bZ_\tN^{\inv}=-\im {\diag}^{-1}(\bx)$ is simply the reciprocal of the diagonal elements.
Hence, by defining 
\begin{equation}
    x_n^\prime = -\frac{R^2}{x_n}
\end{equation}
we arrive at 
    $\bZ_\teR + \bZ_\tN^\prime =\frac{1}{R}\Real{(\bZ_\teR)}^{\frac{1}{2}} (\eyeM R + \im \diag(\bx^\prime)) \Real{(\bZ_\teR)}^{\frac{1}{2}}$
and correspondingly at the new channel [cf. Eqn. \eqref{eq:DecouplingChannelModel}]
\begin{equation}\label{eq:DecoupledChannelGainPowerMatching}
      \bZ^\prime = \bZ_\teDS - \bZ_\teDR^\prime (\eyeM R + \im \diag(\bx^\prime))^{\inv} \bZ_\teRS^\prime
\end{equation}
with the new effective impedance matrices given by
\begin{equation}\label{eq:PowerMatchingNewChannelMatrices}
    \bZ_\teDR^\prime = \bZ_\teDR\Real{(\bZ_\teR)}^{-\frac{1}{2}}\sqrt{R}, \quad \bZ_\teRS^\prime = \sqrt{R}\Real{(\bZ_\teR)}^{-\frac{1}{2}}\bZ_\teRS.
\end{equation}
Analyzing expression \eqref{eq:DecoupledChannelGainPowerMatching}, we can see that the structure is similar to a system without mutual coupling of the \ac{RIS} elements.
Hence, all algorithms and solutions, neglecting the mutual coupling of the \ac{RIS} array, can be extended to mutual coupling by considering the new channel matrices in Eqn. \eqref{eq:PowerMatchingNewChannelMatrices}.

\subsection{Optimization}
The channel gain maximization for a \ac{RIS}-aided \ac{SISO} system without mutual coupling at the \ac{RIS} can be solved in closed-form as has been shown in \cite{MutualCouplingZAlgo}.
Switching to the phase representation \cite{MutualCouplingZAlgo}, \cite{NewChannelModel} the equivalent channel model reads as
\begin{equation}
    z^\prime = z_\teDS + \frac{1}{2R}\bz_\teDR^{\prime,\transpo} (\mThet^\prime - \eyeM)\bz_\teRS^\prime
\end{equation}
where $\mThet^\prime$ is a diagonal matrix with unit-modulus constrained entries on its diagonal.
The optimal solution for channel gain maximization is given by phase alignment (see \cite{MutualCouplingZAlgo}) and we arrive at 
\begin{equation}\label{eq:PowerMatchingClosedForm}
    \abs{z^\prime}^2 = \left(\abs{z_\teDS -\frac{1}{2R}\bz_\teDR^{\prime,\transpo}\bz_\teRS^\prime } + \frac{1}{2R}\summe{n=1}{\NRe}\abs{z_{\teDR,n}^{\prime}}\abs{ z_{\teRS,n}^\prime}\right)^2.
\end{equation}

\begin{figure}[h!]
	\centering
     \includegraphics[scale = 0.8]{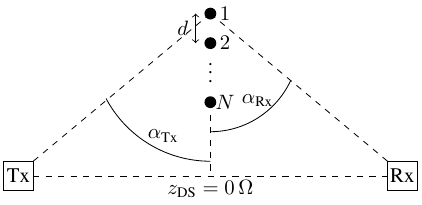}
	\caption{\ac{SISO} Link with $\NRe$ Elements at the \ac{RIS}}
	\label{fig:Scenario}
	
\end{figure}

\subsection{Array Gain Analysis}\label{subsec:ArrayGainAnalysis}
The closed-form solution in Eqn. \eqref{eq:PowerMatchingClosedForm} allows to analyze the array gain analytically.
To this end, we consider a simple \ac{LOS} scenario according to Fig. \ref{fig:Scenario}, where we assume 
\begin{equation}
    z_\teDS = 0 \,\Omega, \quad, \bz_\teDR^\transpo = \sqrt{\gamma_{\teDR}} \ba^\transpo_\teDR R, \quad  \bz_\teRS = \sqrt{\gamma_{\teRS}} \ba_\teRS R
\end{equation}
with the pathlosses $\sqrt{\gamma_{\teDR}}$, $\sqrt{\gamma_{\teRS}}$, and the \ac{LOS} \ac{ULA} vectors $\ba_\teDR = \ba(\alpha_{\text{Rx}}), \; \ba_\teRS = \ba(\alpha_{\text{Tx}})$ where
\begin{equation}
    a_n(\alpha) = e^{-\im(n-1) 2 \pi \frac{d}{\lambda} \cos(\alpha)}.
\end{equation}
For this channel model, the channel gain for one \ac{RIS} element reads as $\abs{z^\prime}^2|_{\NRe =1}=  \frac{{\gamma_{\teDS}}{\gamma_{\teDS}}}{R^2}$.
Normalizing the channel gain for $\NRe$ elements by $\abs{z^\prime}^2|_{\NRe =1}$, we arrive at the array gain
\begin{equation}\label{eq:ArrayGain}
    A = \frac{1}{4} \left(\abs{\ba_\teDR^{\transpo}\bC_\teR^{\inv}\ba_\teRS}+\summe{n=1}{\NRe}\abs{\ba_\teDR^\transpo\bC_\teR^{-\frac{1}{2}}\be_n}\abs{\be^\transpo_n\bC_\teR^{-\frac{1}{2}}\ba_\teRS} \right)^2,
\end{equation}
where we used the additional definition
\begin{equation}
    \bC_\teR = \frac{1}{R} \Real{(\bZ_\teR)}.
\end{equation}
\subsection{No Coupling}
When there is a spacing of $d = \frac{\lambda}{2} k, \; k \in \mathbb{Z}^+$ the matrix $\bC_\teR = \eyeM$ is the identity and there exists no mutual coupling.
It is important to note that this is only the case when deploying a decoupling network.
When there is no decoupling network present, the off-diagonal's imaginary parts of $\bZ_\teR$ are non-zero which the single-connected \ac{RIS} cannot neutralize.
\subsubsection*{Front-Fire}
In case of the front-fire direction we have $\alpha_{\text{Rx}} = \alpha_{\text{Tx}} =\frac{\pi}{2}$ and, hence, $\ba_\teRS = \ba_\teDR = \eye$.
Here, the array gain reads as
\begin{equation}
    \begin{aligned}\label{eq:FrontFireArrayGain}
        A &= \frac{1}{4} \left(\eye^\transpo\bC_\teR^{\inv}\eye+\summe{n=1}{\NRe}\abs{\be^\transpo_n\bC_\teR^{-\frac{1}{2}}\eye}^2 \right)^2\\
        &=\frac{1}{4} \left(\eye^\transpo\bC_\teR^{\inv}\eye+\norm{\bC_\teR^{-\frac{1}{2}}\eye}_2^2 \right)^2\\
        &=  (\eye^\transpo\bC_\teR^{\inv}\eye)^2
    \end{aligned}
\end{equation}
and interestingly, it is exactly the square of a conventional transmit array gain (see \cite{TowardCircuitTheory} for the transmit array gain).

\subsubsection*{End-Fire}
In case of the end-fire direction we have $\alpha_{\text{Rx}}  = \pi,\; \alpha_{\text{Tx}} = 0$ and, hence, $\ba_\teRS = \ba_0,\;\ba_\teDR = \ba_0^*$ with $\ba_0 = \ba(0)$.
Following \eqref{eq:FrontFireArrayGain}, we arrive at the array gain
\begin{equation}
    \begin{aligned}
        A %&= \frac{1}{4} \abs{\ba^\transpo\bC_\teR^{\inv}\ba+\summe{n=1}{\NRe}\abs{\be^\transpo_n\bC_\teR^{-\frac{1}{2}}\ba}^2 }^2\\
        % &=\frac{1}{4}\abs{\ba^\transpo\bC_\teR^{\inv}\ba+\norm{\bC_\teR^{-\frac{1}{2}}\ba}_2^2 }^2\\
        &=\left(\ba_0^\Her \bC_\teR^{\inv}\ba_0 \right)^2 \underset{d \rightarrow 0}{\rightarrow}{\NRe^4}\\    \end{aligned}
\end{equation}
which is again the square of the conventional transmit array gain (see \cite{TowardCircuitTheory}).
Additionally, the term $\ba_0^\Her \bC_\teR^{\inv}\ba_0$ approaches $\NRe^2$ for $d\rightarrow 0$ (see \cite{SquaredArrayGain},\cite{TowardCircuitTheory}).
Hence, we obtain an array gain of $\NRe^4$ for $d \rightarrow 0$.
This super-quadratic gain can only be achieved for lossless antennas.
When incorporating Ohmic losses, we arrive at (see \cite{TowardCircuitTheory})
\begin{equation}
    \bC_\teR^{\text{loss}} = \bC_\teR + \gamma \eyeM
\end{equation}
with $\gamma = \frac{R_\text{d}}{R}$ where $R_\text{d}$ is the dissipation resistance. 
Here, the array gain decreases with an incresing $\gamma$.
The performance will be analyzed in the next section.
\section{Numerical Results}\label{sec:NumericalResults}
In this section, we evaluate the decoupling network together with the mutual coupling aware algorithms in the same \ac{LOS} scenario as in Section \ref{subsec:ArrayGainAnalysis} according to Fig. \ref{fig:Scenario}.

\begin{figure}[!t]%[!t]
    \includegraphics{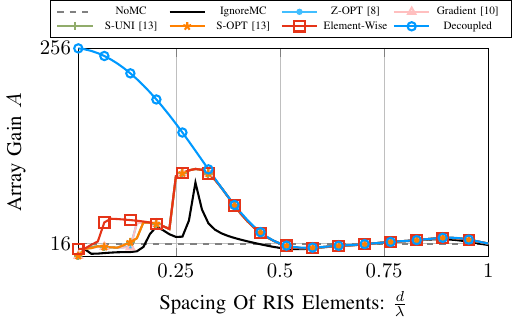}
    \vspace*{-0.3cm}
	\caption{Method Comparison for $\NRe = 4$ and end-fire direction}
\label{fig:ComparisonAlgorithms}
\end{figure}

\begin{figure}[!b]
    \includegraphics{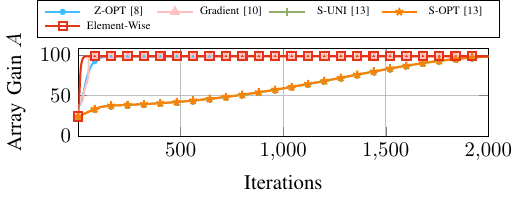}
    \vspace*{-0.3cm}
	\caption{Method Comparison for $\NRe = 4$ and end-fire direction with $d= \frac{\lambda}{4}$}
    \label{fig:ConvergenceAlgorithms}
\end{figure}

\subsection{Algorithm Comparison}
We start by comparing the decoupling network approach (Decoupled) together with mutual coupling aware algorithms for $\NRe=4$ elements.
The end-fire direction is chosen as, for $d\rightarrow 0$, the matrix $\bZ_\teR$ becomes singular resulting in a numerically challenging scenario in which mutual coupling plays a significant role.
We compare the decoupling network with the element-wise approach discussed in this article (Element-Wise) together with the gradient approach of \cite{MutualCouplingGradient} (Gradient), the Neumann series approach of \cite{MutualCouplingZAlgo} (Z-OPT), and the two scattering parameter based approaches of \cite{FollowUPSZJournal}, \cite{FollowUpSZ} (S-UNI and S-OPT).
All algorithms have a complexity of $\mathcal{O}(\NRe^3)$ per iteration except S-OPT which needs to solve another optimization problem in each of the iterations (see \cite{FollowUPSZJournal} for an efficient implementation based on the bisection method).
To futher evaluate the results, we include a purely theoretical model without mutual coupling (NoCoupling) where we set $\bZ_\teR = \eyeM R$ and a primitive optimization (IgnoreMC) where we use the solution of NoCoupling (which is $\mThet_{\text{NoMC}}^\prime = -\eyeM$) but then use the actual model with the non-diagonal coupling matrix $\bZ_\teR$ for the evaluation.
All the methods above are also initialized with the solution of NoCoupling $\mThet_{\text{NoMC}}^\prime = -\eyeM$.
The metric is the array gain [see Eqn. \eqref{eq:ArrayGain}], which is the normalized channel gain.

In Fig. \ref{fig:ComparisonAlgorithms}, we compare all the different methods discussed above.
Decoupled is clearly outperforming all mutual coupling algorithms as it finds the optimal solution in closed-form.
The mutual coupling aware algorithms are mostly overlapping, only the Element-Wise approach is performing slightly better for small spacings.
Additionally, this method has the advantage that it not based on an approximation as it is the case for the Neumann series based approaches S-OPT, S-UNI, and Z-OPT.

Moreoever, it appears in Fig. \ref{fig:ComparisonAlgorithms} that the mutual coupling algorithms become numerically unstable as $d\rightarrow 0$.
However, this is not the case and they only converge to suboptimal local maxima.
This is illustrated in Fig. \ref{fig:ConvergenceAlgorithms} for $d=\frac{\lambda}{4}$ where we can see that all algorithms are converging.
Nevertheless, all of them are stuck in the same suboptimal local maximum.
The Decoupled method is able to circumvent this situation as it obtains the optimal solution in closed-form.
In Fig. \ref{fig:ConvergenceAlgorithms}, we can see that S-OPT and S-UNI require a lot of iterations to converge.
However, this is only the case for this specific scenario and also different observations have been made (see \cite{FollowUPSZJournal}).

\begin{figure}[!t]%[!t]
	\centering
    \includegraphics{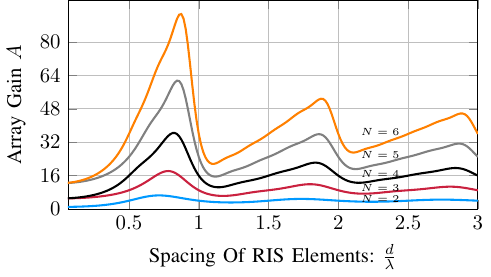}
    \vspace*{-0.3cm}
	\caption{Array Gain for the front-fire direction of the array}
\label{fig:FrontFire}
% \vspace*{-0.6cm}
\end{figure}

\begin{figure}[!b]%[!t]

	\centering
    \includegraphics{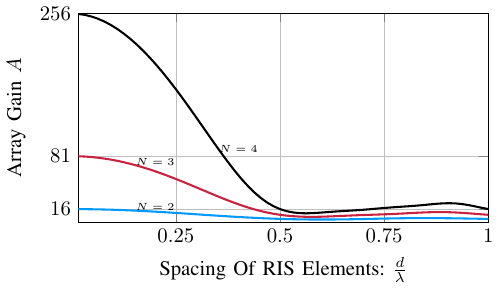}
    \vspace*{-0.3cm}
	\caption{Array Gain for the end-fire direction of the array}
\label{fig:EndFire}
% \vspace*{-0.6cm}
\end{figure}

\begin{figure}[!t]%[!t]

	\centering
    \includegraphics{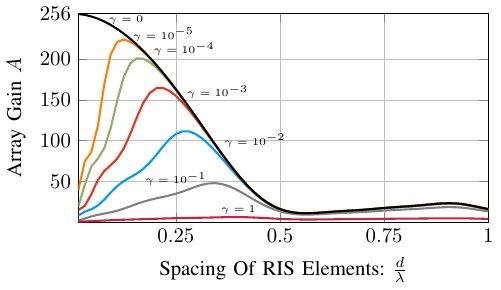}
    \vspace*{-0.3cm}
	\caption{Array Gain for the end-fire direction of the array for $\NRe=4$}
\label{fig:EndFireLossy}
% \vspace*{-0.6cm}
\end{figure}

\begin{figure}[!h]%[!t]
	\centering
    \includegraphics{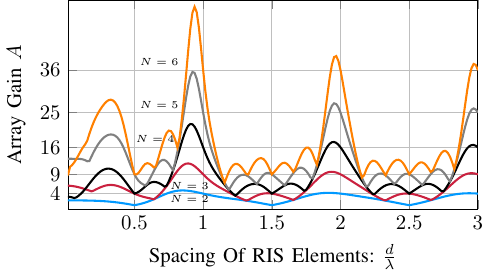}
    \vspace*{-0.3cm}
	\caption{Array Gain for $\alpha_{\text{Tx}} =\frac{\pi}{2}$ and $\alpha_{\text{Rx}} = 0$}
\label{fig:Corner}
%  \vspace*{-1cm}
\end{figure}

\subsection{Analysis of the Decoupling Network}
We have seen from the last subsection, that the decoupling network (Decoupled) with its optimal solution given in closed form shows the best performance for handling mutual coupling in the \ac{RIS} antenna array.
Additionally, this closed-form solution allows a simple analysis of the performance and we will discuss the potential of decoupling networks in the following.
Starting with the front-fire direction in Fig. \ref{fig:FrontFire}, we obtain \cite[Fig. 6]{TowardCircuitTheory} with squared values.
Therefore, also here, the interesting fact that for $2\NRe$ and $2\NRe-1$ elements, the array gains converge to the same value as $d \rightarrow 0$ can be observed.

Analyzing the end-fire direction in Fig. \ref{fig:EndFire}, we obtain \cite[Fig. 5]{TowardCircuitTheory} with squared values and, hence,
$\NRe^4$ is approached as $d\rightarrow 0$. However, as discussed in Section \ref{subsec:ArrayGainAnalysis}, this super-quadratic gain can only be obtained for a lossless \ac{RIS}.
When considering heatloss in Fig. \ref{fig:EndFireLossy} for $\NRe = 4$ (which is \cite[Fig. 11]{TowardCircuitTheory} with squared values), the gain decreases as $d \rightarrow 0$.
Nevertheless, a considerable gain can be achieved for a sub half-wavelength spacing.

\begin{figure}[!ht]%[!t]

	\centering
    \includegraphics{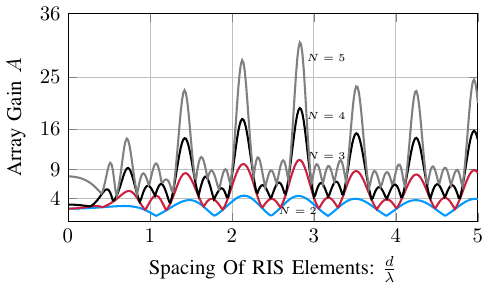}

    \vspace*{-0.3cm}
    \caption{Array Gain for $\alpha_{\text{Tx}} =\frac{\pi}{4}$ and $\alpha_{\text{Rx}} = \frac{\pi}{4}$}
    \label{fig:SpacingInOut}
 \vspace*{-0.6cm}

\end{figure}

For other geometries, i.e.,  $\alpha_{\text{Tx}} =\frac{\pi}{2}$ and $\alpha_{\text{Rx}} = 0$, or $\alpha_{\text{Tx}} =\frac{\pi}{4}$ and $\alpha_{\text{Rx}} = \frac{\pi}{4}$,
the gain appears to be lower as shown in Figs. \ref{fig:Corner} and \ref{fig:SpacingInOut}.
Especially in Fig. \ref{fig:Corner}, there is the interesting aspect that $2\NRe -1$ elements provide a larger gain than $2\NRe$ elements when $d \rightarrow 0$.
Overall, we can observe that the array gain of the \ac{RIS} when including mutual coupling is clearly different for different scenarios,
where the end-fire direction provides the largest gains.

\section{Conclusion}
The new approach of the Element-Wise algorithm provides a computationally efficient algorithm which achieves at least the same performance as existing approaches.
However, all mutual coupling aware algorithms are not guaranteed to converge to the global optimum which leads to a performance degradation.
On the other hand, the proposed decoupling networks lead to a simplification of the system model resembling the structure of the case without mutual coupling.
Hence, a closed-form solution is possible for the \ac{SISO} channel gain maximization and the decoupling networks are clearly outperforming all mutual coupling aware algorithms w.r.t. computational complexity as well as the performance.
However, even though the decoupling networks have excellent performance, there is a significant hardware cost as the number of elements in the decoupling networks scales quadratically with the number of reflecting elements.
Therefore, future works will especially focus on partially connected networks which only scale linearly with the number of reflecting elements.

% \newpage

\bibliographystyle{IEEEtran}
\bibliography{refs}

% Generated by IEEEtran.bst, version: 1.14 (2015/08/26)
\begin{thebibliography}{10}
\providecommand{\url}[1]{#1}
\csname url@samestyle\endcsname
\providecommand{\newblock}{\relax}
\providecommand{\bibinfo}[2]{#2}
\providecommand{\BIBentrySTDinterwordspacing}{\spaceskip=0pt\relax}
\providecommand{\BIBentryALTinterwordstretchfactor}{4}
\providecommand{\BIBentryALTinterwordspacing}{\spaceskip=\fontdimen2\font plus
\BIBentryALTinterwordstretchfactor\fontdimen3\font minus
  \fontdimen4\font\relax}
\providecommand{\BIBforeignlanguage}[2]{{%
\expandafter\ifx\csname l@#1\endcsname\relax
\typeout{** WARNING: IEEEtran.bst: No hyphenation pattern has been}%
\typeout{** loaded for the language `#1'. Using the pattern for}%
\typeout{** the default language instead.}%
\else
\language=\csname l@#1\endcsname
\fi
#2}}
\providecommand{\BIBdecl}{\relax}
\BIBdecl

\bibitem{Power_Min_IRS}
Q.~Wu and R.~Zhang, ``{I}ntelligent {R}eflecting {S}urface {E}nhanced
  {W}ireless {N}etwork via {J}oint {A}ctive and {P}assive {B}eamforming,''
  \emph{IEEE Transactions on Wireless Communications}, vol.~18, no.~11, pp.
  5394--5409, 2019.

\bibitem{SmartRadioEnvironment}
M.~Di~Renzo, A.~Zappone, M.~Debbah, M.-S. Alouini, C.~Yuen, J.~de~Rosny, and
  S.~Tretyakov, ``Smart radio environments empowered by reconfigurable
  intelligent surfaces: How it works, state of research, and the road ahead,''
  \emph{IEEE Journal on Selected Areas in Communications}, vol.~38, no.~11, pp.
  2450--2525, 2020.

\bibitem{EnergyEff}
C.~Huang, A.~Zappone, G.~C. Alexandropoulos, M.~Debbah, and C.~Yuen,
  ``{R}econfigurable {I}ntelligent {S}urfaces for {E}nergy {E}fficiency in
  {W}ireless {C}ommunication,'' \emph{IEEE Transactions on Wireless
  Communications}, vol.~18, no.~8, pp. 4157--4170, 2019.

\bibitem{MutualCouplingAware}
G.~Gradoni and M.~Di~Renzo, ``End-to-end mutual coupling aware communication
  model for reconfigurable intelligent surfaces: An electromagnetic-compliant
  approach based on mutual impedances,'' \emph{IEEE Wireless Communications
  Letters}, vol.~10, no.~5, pp. 938--942, 2021.

\bibitem{TowardCircuitTheory}
M.~T. Ivrlač and J.~A. Nossek, ``Toward a circuit theory of communication,''
  \emph{IEEE Transactions on Circuits and Systems I: Regular Papers}, vol.~57,
  no.~7, pp. 1663--1683, 2010.

\bibitem{ScatteringRIS}
S.~Shen, B.~Clerckx, and R.~Murch, ``Modeling and architecture design of
  reconfigurable intelligent surfaces using scattering parameter network
  analysis,'' \emph{IEEE Transactions on Wireless Communications}, vol.~21,
  no.~2, pp. 1229--1243, 2022.

\bibitem{NewChannelModel}
J.~A. Nossek, D.~Semmler, M.~Joham, and W.~Utschick, ``Physically consistent
  modelling of wireless links with reconfigurable intelligent surfaces using
  multiport network analysis,'' \emph{arXiv 2308.12223}, 2023.

\bibitem{MutualCouplingZAlgo}
X.~Qian and M.~D. Renzo, ``{M}utual {C}oupling and {U}nit {C}ell {A}ware
  {O}ptimization for {R}econfigurable {I}ntelligent {S}urfaces,'' \emph{IEEE
  Wireless Communications Letters}, vol.~10, no.~6, pp. 1183--1187, 2021.

\bibitem{MutualCouplingSumRate}
A.~Abrardo, D.~Dardari, M.~Di~Renzo, and X.~Qian, ``{MIMO} {I}nterference
  {C}hannels {A}ssisted by {R}econfigurable {I}ntelligent {S}urfaces: {M}utual
  {C}oupling {A}ware {S}um-{R}ate {O}ptimization {B}ased on a {M}utual
  {I}mpedance {C}hannel {M}odel,'' \emph{IEEE Wireless Communications Letters},
  vol.~10, no.~12, pp. 2624--2628, 2021.

\bibitem{MutualCouplingGradient}
M.~Akrout, F.~Bellili, A.~Mezghani, and J.~A. Nossek, ``{P}hysically
  {C}onsistent {M}odels for {I}ntelligent {R}eflective {S}urface-assisted
  {C}ommunications under {M}utual {C}oupling and {E}lement {S}ize
  {C}onstraint,'' \emph{arXiv 2302.11130}, 2023.

\bibitem{FollowUpBDRIS}
H.~Li, S.~Shen, M.~Nerini, M.~D. Renzo, and B.~Clerckx, ``Beyond diagonal
  reconfigurable intelligent surfaces with mutual coupling: Modeling and
  optimization,'' \emph{arXiv 2310.02708}, 2024.

\bibitem{FollowUpSZ}
A.~Abrardo, A.~Toccafondi, and M.~D. Renzo, ``Analysis and optimization of
  reconfigurable intelligent surfaces based on $s$-parameters multiport network
  theory,'' \emph{arXiv 2308.16856}, 2023.

\bibitem{FollowUPSZJournal}
------, ``Design of reconfigurable intelligent surfaces by using s-parameter
  multiport network theory -- optimization and full-wave validation,''
  \emph{arXiv 2311.06648}, 2023.

\bibitem{ElementWise}
H.~E. Hassani, X.~Qian, S.~Jeong, N.~S. Perović, M.~D. Renzo, P.~Mursia,
  V.~Sciancalepore, and X.~Costa-Pérez, ``Optimization of ris-aided mimo -- a
  mutually coupled loaded wire dipole model,'' \emph{arXiv 2306.09480}, 2023.

\bibitem{SquaredArrayGain}
E.~Altshuler, T.~O'Donnell, A.~Yaghjian, and S.~Best, ``{A} monopole
  superdirective array,'' \emph{IEEE Transactions on Antennas and Propagation},
  vol.~53, no.~8, pp. 2653--2661, 2005.

\end{thebibliography}

\end{document}